\documentstyle[psfig,float,twocolumn,eqsecnum,aps]{revtex}

\psfigurepath{/net/del/turu/natali/couet/my-res}

\def\Om{\it \Omega}
\def\q{\qquad}
\def\beg{\begin{eqnarray}}
\def\ende{\end{eqnarray}}
\def\gsim{\lower.4ex\hbox{$\;\buildrel >\over{\scriptstyle\sim}\;$}} 
\def\lsim{\lower.4ex\hbox{$\;\buildrel <\over{\scriptstyle\sim}\;$}}

\newcommand{\Pm}{\mbox{Pm}}
\begin{document}
\title{The stability of  axisymmetric  Taylor-Couette flow in hydromagnetics}
\author{G\"unther R\"udiger}
\address{Astrophysikalisches Institut Potsdam,
         An der Sternwarte 16, D-14482 Potsdam, Germany}
 \author{Dima Shalybkov}      
 \address{A.F. Ioffe Institute for Physics and Technology,
          194021, St. Petersburg, Russia}
\date{\today}
\maketitle

\begin{abstract}
The linear marginal instability of an  axisymmetric   MHD Taylor-Couette flow
of infinite vertical extension is considered. Only those vertical wave numbers
are interesting for which the eigenvalue the (`characteristic') Reynolds number
is minimal. For flows with a resting outer cylinder there is a well-known 
characteristic Reynolds number even without magnetic field but for certain 
(weak) magnetic fields there are solutions with {\em smaller} Reynolds numbers
so that a characteristic minimum exists.     We call those Reynolds numbers,
the related wave number and the related Hartmann number as their  {\em critical}
 values.  The minimum only exists, however,  for not too small magnetic Prandtl
numbers 
 (see Figs. \ref{fig} and \ref{fig0} for a typical example).
For small magnetic Prandtl numbers -- or sufficiently small gaps -- one only finds
the typical magnetic-originated suppression of any instability.

 More  interesting  are   experiments  
where the outer cylinder rotates so fast that the Rayleigh 
criterion for  hydrodynamic  {\em stability} is fulfilled. 
We find that for given geometry and given magnetic Prandtl number now
always a magnetic field amplitude exists where the characteristic
Reynolds number is minimal.  These critical values are  computed for
different magnetic Prandtl numbers and for three types of geometry
(small, medium and wide gaps between the rotating cylinders). In all cases 
the Reynolds numbers are running with 1/Pm
for small enough Pm, and the critical Reynolds numbers exceed values of 10$^6$
for the  magnetic Prandtl number of sodium (10$^{-5}$) 
or gallium (10$^{-6}$).

The container walls are considered as either electrically conducting or as 
isolators. Compared with the results for conducting walls, for small
and medium size gaps between the cylinders i) the critical Reynolds number
is smaller, ii) the critical Hartmann number is higher and iii) the  Taylor
vortices are longer in the vertical direction for  isolating walls.
For experiments with wide gaps the differences between both sets of boundary
conditions become smaller and smaller. 
\end{abstract}


\pacs{Valid PACS}

\section{Introduction}
The longstanding problem of the generation of turbulence in various
hydrodynamically stable situations has found a solution in recent years
with the so called `Balbus-Hawley instability', in
which the presence of a magnetic field has a destabilizing effect on a
differentially rotating flow, provided that the angular velocity decreases
outwards with the radius, \cite{BH91}.
This instability has been  discovered decades ago \cite{V59}, \cite{C61} for
ideal Couette flow, but only after Balbus and Hawley it was recognized the 
importance of this magnetorotational instability (MRI) as the source of
turbulence in the accretion discs with differential (Keplerian) rotation.

However, the MRI has never been observed in the laboratory (\cite{DO60},
\cite{DO62}, \cite{DC64}, \cite{B70}). Moreover, Chandrasekhar \cite{C61}
already suggested  the existence of MRI for ideal Taylor-Couette flow, but
his results for non-ideal fluids for small gaps and within the  small
magnetic Prandtl number approximation  demonstrated the absence of MRI for
hydrodynamically stable flow. 
Recently, Goodman and Ji \cite{GJ01} claimed that this absence of MRI
was due to the use of the small magnetic Prandtl number limit. 
The magnetic Prandtl number is really very small under laboratory conditions
($\sim 10^{-5}$ and smaller).  Obviously, the understanding of this phenomenon
is very important for possible future experiments, Taylor-Couette flow  dynamo
experiments included.
\begin{figure}
\psfig{figure=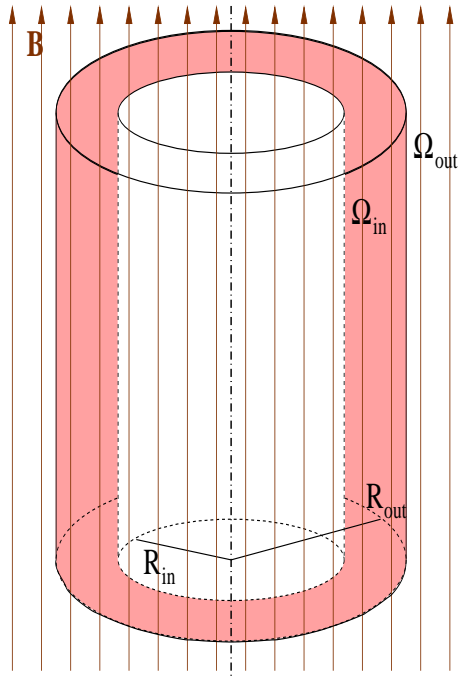,width=6cm,height=9cm}
\caption{Cylinder geometry of the  Taylor-Couette flow.}
\label{geometry}
\end{figure}
Here,  the dependence of real
Couette flow on magnetic Prandtl number and gap-width between rotating
cylinders is investigated.  The simple model of uniform  density fluid
contained between two vertically-infinite rotating cylinders is used with 
constant magnetic field parallel to the rotation axis. The unperturbed state
is any stationary circular flows of an incompressible fluid. 
In the absence of viscosity, the class of such  flows is 
very wide: indeed,
if $\Om$ denotes the angular velocity of rotation about the axis, then 
the equations of motion allow $\Om$ to be an arbitrary function of the 
distance $R$ from the axis, provided the velocities in the radial and
the axial directions are zero.
For viscous flows, however,  the class
becomes very restricted: in fact, in the absence of any transverse 
pressure gradient, the most general form of $\Om$  allowed is
\begin{equation}
\Om(r) = a+b/{R}^2,
\label{Om}
\end{equation}
where $a$ and $b$ are two constants related to the angular
velocities $\Om_{\rm in}$ and $\Om_{\rm out}$ with which the inner
and the outer cylinders are rotating. If $R_{\rm in}$ and $R_{\rm out}$
($R_{\rm out}>R_{\rm in}$) are the radii of the two cylinders then
\begin{equation}
a=\Om_{\rm in}{\hat \mu-{\hat\eta}^2\over1-{\hat\eta}^2}
\; \; {\rm and} \; \;
b=\Om_{\rm in} R_{\rm in}^2 {(1-\hat\mu)\over1-{\hat\eta}^2},
\label{ab}
\end{equation}
with 
\begin{equation}
\hat\mu=\Om_{\rm out}/\Om_{\rm in}  \q  {\rm and} \q 
\hat\eta=R_{\rm in}/R_{\rm out}.
\label{mu}
\end{equation}
After the Rayleigh stability criterion, $d(R^2 \Omega)^2/dR>0$,
rotation laws are hydrodynamically stable
for $\hat\mu>\hat\eta^2$. Taylor-Couette flows with resting outer cylinders 
($\hat\mu=0$) are thus never stable. 

Here, in order to isolate the MRI we 
are mainly interested in flows with rotating outer cylinders  so that the
hydrodynamic-stability criterion   $\hat\mu>\hat\eta^2$ is fulfilled.
Our standard example is formed with $\hat\eta$=0.5 and $\hat\mu=0.33$.
\section{Basic equations}
$R$, $\phi$, and $z$ are the cylindrical coordinates.  A viscous 
electrically-conducting incompressible fluid between
two rotating infinite cylinders in the presence of a uniform magnetic
field  parallel to the rotation axis leads to the basic solution
$U_R=U_z=B_R=B_\phi=0$
\beg
B_z=B_0={\rm const.} \q  U_\phi=aR+b/R, 
\ende
with  $U_i$ as  the velocity  and  $B_i$ the magnetic field, 
$a$ and $b$ are given by (\ref{ab}).  We are interested in the stability of
this solution. The perturbed state of the flow may be  described by
\beg
u_R, \; U_\phi+u_\phi, \; u_z, \; b_R, \; b_\phi, \; B_0+b_z,
\; \delta P,
\ende
with  $\delta P$ as the  pressure perturbation.

Here only  the linear stability problem with axisymmetric perturbations
is considered.
By analyzing the disturbances into normal modes  the solutions
of the linearized magnetohydrodynamical equations are of the form
\begin{eqnarray}
&&u_R=u_R(R)e^{\omega t} \cos (kz), 
\ \ \  b_R=b_R(R)e^{\omega t} \sin (kz),  \nonumber \\
&&u_\phi=u_\phi(R)e^{\omega t} \cos (kz),
\ \ \ \   b_\phi=b_\phi(R)e^{\omega t} \sin (kz), \nonumber \\
&&u_z=u_z(R)e^{\omega t} \sin (kz),
\ \ \ \ \  b_z=b_z(R)e^{\omega t} \cos (kz).
\end{eqnarray}
 Stationary modes are always more critical than oscillatory ones,
according to the results in \cite{C61} and \cite{CS67}. So,
only marginal stability  will be considered ($\omega =0$).
The derivation of the equations describing this situation is due to  
Chandrasekhar \cite{C61}; it should not to be   repeated here.
We only differ in the  normalizations. Let  $d=R_{\rm out} - R_{\rm in}$
be the gap between
the cylinders. We use 
\begin{equation}
H=(R_{\rm in}d)^{1/2}
\label{2.4}
\end{equation}
 as unit of length,
the Alfv\'{e}n velocity $V_{\rm A}=B_0/(\mu_0 \rho)^{1/2}$ as unit of
perturbed velocity and $B_0 \cdot\Pm^{1/2}$ as unit of perturbed
magnetic field with the magnetic Prandtl number 
\begin{equation}
{\rm Pm} = {\nu\over\eta},
\label{pm}
\end{equation}
 $\nu$ is the
kinematic viscosity, $\eta$ is the magnetic diffusivity. 
Note  $H^{-1}$ as the unit of wave numbers.

Using the same symbols for normalized quantities as before the  equations
take the form
\begin{eqnarray}
&&(DD_*-k^2)^2u_R+k^2{\rm Ha}^2u_R-2k^2{\rm Re}
{\Om \over \Om_{\rm in}}u_\phi =0, \nonumber \\
&&(DD_*-k^2)u_\phi+k {\rm Ha} b_\phi - {\rm Re}
{1 \over R} {d \over dR} \left( R^2 {\Om \over \Om_{\rm in} }
\right) u_R=0, \nonumber \\
&&(DD_*-k^2)b_R-k {\rm Ha} u_R=0, \nonumber \\
&&(DD_*-k^2)b_\phi-k {\rm Ha} u_\phi + {\rm Re}
\Pm R {d \over dR} \left( {\Om \over \Om_{\rm in} } \right) b_R=0\nonumber\\
\label{syst}
\ende
with
\beg
{\rm Ha}={B H \over \sqrt{\mu_0 \rho \nu \eta}} \q 
{\rm Re}={\Om_{\rm in} H^2 \over \nu},
\label{numb}
\ende
where ${\rm Ha}$ is the Hartmann number, ${\rm Re}$ is the Reynolds
number of the inner rotation, $\rho$ is the
density, $\mu_0$ is the magnetic constant.  Chandrasekhar's  
notations $D=d/dR$ and $D_*=d/dR + 1/R$ are also used.

\section{Boundary conditions}

An appropriate set of ten boundary conditions is needed to solve  the system
(\ref{syst}). The situation is more difficult than in the
small-gap-small-Prandtl-number case where only eight boundary
conditions are needed.  Always no-slip conditions for the velocity on the walls
are used, i.e. 
\beg
u_R=0, \q u_\phi=0, \q {d u_R \over dR}=0.
\label{bvel}  
\ende
(see \cite{C61}).
The magnetic boundary conditions depend on the electrical properties
of the walls. The transverse currents and perpendicular component of
magnetic field should vanish on conducting walls, hence
\beg
{d b_\phi \over dR} + {b_\phi \over R}=0, \q b_R=0.
\label{cond}
\ende
The above boundary conditions (\ref{bvel}) and (\ref{cond}) are valid 
for $R=R_{\rm in}$ and  for $R=R_{\rm out}$. 

The situation changes for
isolating  walls.
The magnetic field must  match the external magnetic field for
nonconducting walls. The  boundary conditions are different at $R=R_{\rm in}$
and $R=R_{\rm out}$ due to the different behaviour of the modified Bessel
functions for  $R\to 0$ and $R\to \infty$, i.e.  
\beg
b_\phi=0, \; \;
{\partial \over \partial R} (R b_R)=b_R {k R I_0(kR)
\over I_1(kR) } \; \; {\rm for} \; \; R=R_{\rm in},
\label{nonin}
\ende
and
\begin{eqnarray}
&&b_\phi=0, \; \;
{\partial \over \partial R} (R b_R)=-b_R {k R K_0(kR)
\over K_1(kR) } \; \; {\rm for} \; \; R=R_{\rm out},\nonumber\\
\label{nonout}
\end{eqnarray}
 where $I_n$ and $K_n$ are the modified Bessel functions (cf. \cite{GJ01}).

For a fixed Hartmann number, a fixed Prandtl number and a given vertical wave
number we find the eigenvalues of the equation system. They are always minimal 
for a certain wave number which by itself defines the marginally unstable mode.
The corresponding eigenvalue is the desired Reynolds number. 
\section{Results for conducting walls}
We start with the results for  containers  with conducting walls and resting 
outer cylinders but with various gap sizes (medium,  wide and small).  In all 
these cases there are linear instabilities even without magnetic fields. The 
influence of the magnetic field is the question.  If the resulting eigenvalue 
with magnetic field exceeds the eigenvalue without magnetic field then we have 
only the well-known effect of magnetic stabilization  rather than magnetic 
destabilization. As we shall see,  this is indeed the case for sufficiently 
small magnetic Prandtl numbers and/or for containers with a
 a small gap between the cylinders.

\subsection{Resting outer cylinder}
In Fig. \ref{fig}  a resting outer cylinder is considered ($\hat\mu$=0)
for a medium-size gap of $\hat\eta$=0.5 
and for Pm=1. As we know for vanishing magnetic field and  for $\hat\eta$=0.5 
the exact Reynolds number for this case is  about 68 -- well represented by 
the result for Ha=0  in Fig. \ref{fig}. 
But for increasing magnetic field  the Reynolds number is reduced so that figure
the excitation of the Taylor vortices becomes easier 
than without magnetic field. The minimum Reynolds number Re$_{\rm crit}$ of
about 63 for Pm=1
is reached for Ha$_{\rm crit}\simeq$ 4...5.  This  magnetic induced subcritical
excitation of Taylor vortices is due to the MRI.
Always for a (say) critical Hartmann number the Reynolds numbers 
take a minimum which we shall call the critical Reynolds number.
For even stronger 
magnetic fields -- as it must be -- the magnetic field starts to suppress 
the instability so that the Reynolds number starts to grow to infinity.
\begin{figure}
\psfig{figure=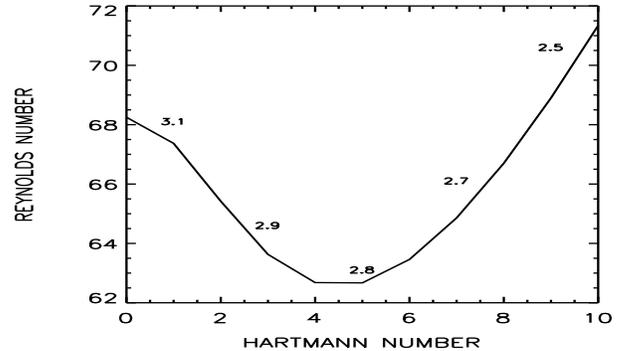,width=8.8cm,height=5.0cm} 
\caption{The  stability line
 for Taylor-Couette flow with resting
 outer cylinder for $\hat\eta=0.5$ and  Pm$=1$.
  The flow is unstable above the line. There is instability even 
 without magnetic fields but  its excitation is easier with magnetic
 fields with Ha $\simeq 4.5$. The  line is marked with those  
     wave numbers for which the eigenvalues are minimal. } 
\label{fig}
\end{figure}
In Fig. \ref{fig0}  the same container is considered but for the small 
magnetic Prandtl number of $10^{-5}$.
 The minimum  characteristic for Pm=1 completely disappears, only  
 suppression of the instability by the magnetic field can be observed.
\begin{figure}
\psfig{figure=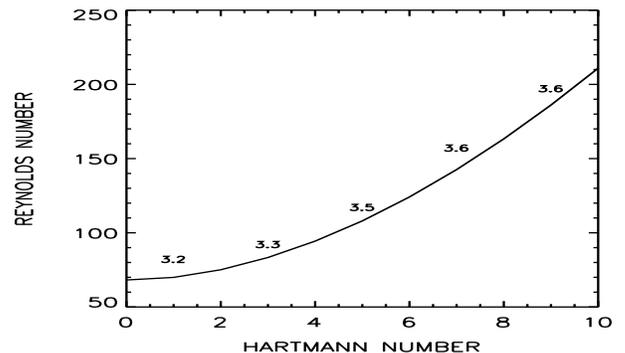,width=8.8cm,height=5.0cm} 
\caption{The  same as in Fig. \ref{fig} but   for  Pm=$10^{-5}$.  The 
minimum characteristic for Pm=1 completely disappears.} 
\label{fig0}
\end{figure}
A container with a small gap ($\hat\eta=0.95$) between the two cylinders 
is now under consideration (Figs. \ref{fig10} and \ref{fig11}). 
Only magnetic suppression of the Taylor-Couette flow instability is 
observed in this case. This is the reason why Chandrasekhar did not find 
the MRI by his detailed numerical simulations for small gaps and very small
magnetic Prandtl numbers. Fig. \ref{fig11} are representing the 
small-gap-small-Prandtl approximation used by Chandrasekhar \cite{C61}. 
In order to find a minimum due to the MRI the magnetic Prandtl number must 
exceed 1, e.g. for Pm$=10$. The smaller the gap, the larger the Pm must be.
\begin{figure}
\psfig{figure=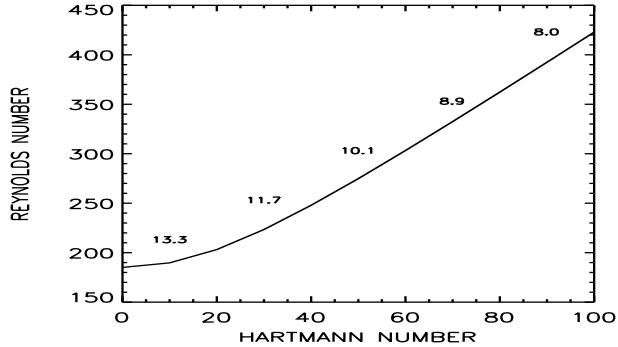,width=8.8cm,height=5.0cm} 
\caption{The stability line
for the flow in a 
small gap ($\hat\eta$=0.95) with resting outer cylinder and for Pm$=1$.  Note 
the disappearance of any minimum  of the Reynolds number.} 
\label{fig10}
\end{figure}
\begin{figure}
\psfig{figure=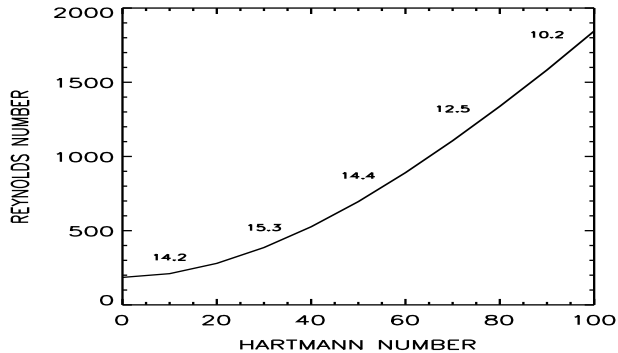,width=8.8cm,height=5.0cm} 
\caption{The same as in Fig. \ref{fig10} but for Pm=$10^{-5}$. 
    } 
\label{fig11}
\end{figure}
In Fig. \ref{fig7}  the results for a container with a wide gap between 
the cylinders are given.
Again we find the magnetic field only suppressing the instability
 for small magnetic Prandtl number. 
Obviously, the MRI does not work efficiently in the limit of small magnetic
Prandtl numbers, i.e. for too low electrical conductivity
Thus, if the electrical conductivity is so small as it is for sodium
or gallium then the MRI cannot be observed by corresponding experiments
with hydrodynamically unstable flows.
\begin{figure}
\psfig{figure=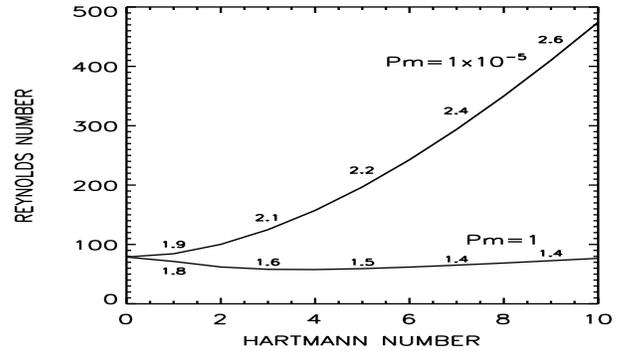,width=8.8cm,height=5.0cm}
\caption{The stability line for the flow in a 
     wide gap ($\hat\eta=0.25$) with resting outer cylinder ($\hat \mu =0$) for
     Pm=1 and Pm$=10^{-5}$.} 
\label{fig7}
\end{figure}
\subsection{Rotating outer cylinder}
Another situation holds if  the outer cylinder  may rotate so fast that the
rotation law does not longer fulfill the Rayleigh criterion and a solution 
for Ha=0  cannot exist. Then the nonmagnetic eigenvalue along the vertical 
axis moves to infinity and we should always have a minimum.   It is the basic
situation in astrophysical applications such for accretion disks with a Kepler
rotation law. Here in this paper the question is whether the critical Reynolds
number and the critical Hartmann number can experimentally be realized. 
The Figs. \ref{fig2}...\ref{fig4}  present the results for both various
Hartmann numbers and magnetic Prandtl 
numbers for a medium-sized gap of $\hat\eta$=0.5. 
\begin{figure}
\psfig{figure=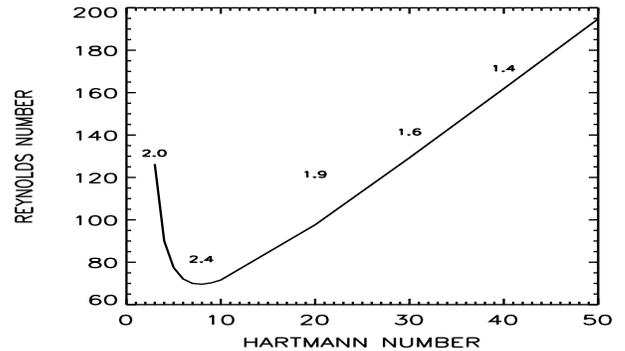,width=8.8cm,height=5.0cm}
     \caption{ The stability line
      for $\hat\eta=0.5$ and
      Pm=1. The outer cylinder  rotates with 33\% of the rotation rate of the
      inner cylinder so that after the Rayleigh criterion the hydrodynamic
      instability for Ha=0 disappears. The minimal Reynolds number is almost 
      the same as in Fig. \ref{fig}.} 
\label{fig2}
\end{figure}
\begin{figure}
\psfig{figure=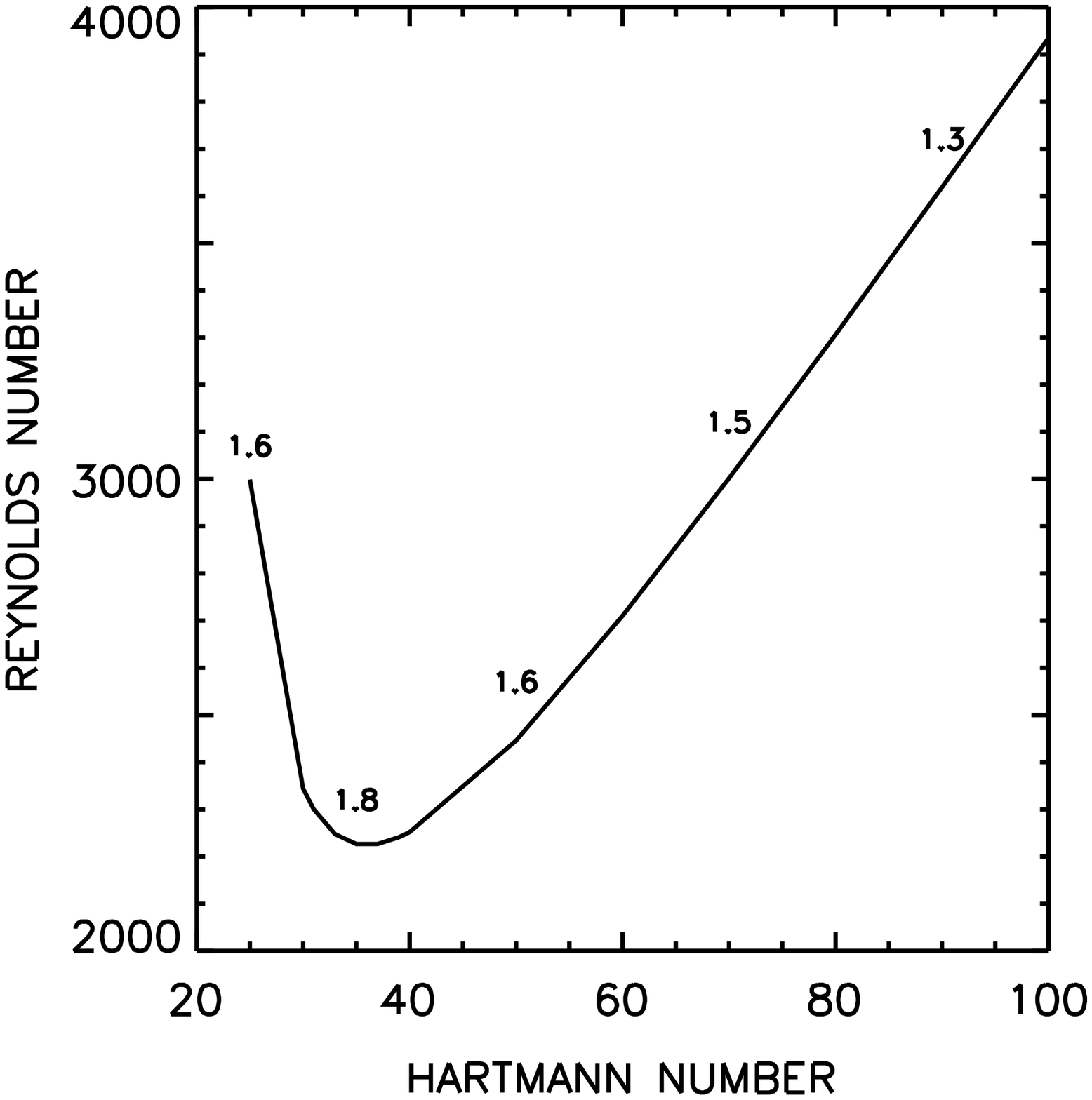,width=8.8cm,height=5.0cm} 
     \caption{The same as in Fig. \ref{fig2} but for  Pm$=10^{-2}$.} 
      \label{fig3}
\end{figure}
\begin{figure}
\psfig{figure=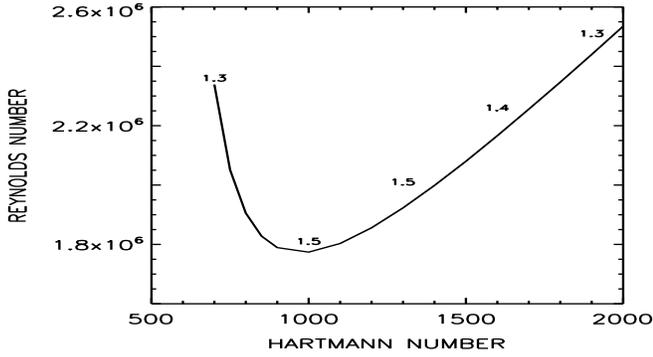,width=8.8cm,height=5.0cm} 
     \caption{The same as in Fig. \ref{fig2} but for Pm=$10^{-5}$.} 
\label{fig4}
\end{figure}
There are always minima of the characteristic Reynolds numbers for 
certain Hartmann numbers. The minima and the critical Hartmann numbers 
increase for decreasing magnetic Prandtl numbers.
For $\hat\eta$=0.5 and $\hat\mu$=0.33 the  critical Reynolds numbers together 
with the critical Hartmann numbers are plotted in Fig. \ref{fig5}. 

For  the small magnetic Prandtl numbers we find interesting and simple 
relations. With
\beg
C_\Omega = {\rm Re} {\rm Pm}
\label{COm}
\ende
and
\beg
{\rm Ha}^* = {\rm Ha} \sqrt{\rm Pm}
\label{Ha-st}
\ende
it follows
\beg
C_\Omega \simeq 20
\label{COmega}
\ende
and
\beg
{\rm Ha}^* \simeq 3.5.
\label{HA}
\ende
$C_\Omega$ is the magnetic Reynolds number, $C_\Omega = \Om_{\rm in} 
H^2/\eta$ 
(or dynamo number) and Ha$^*$ is the magnetic Hartmann number Ha$^* = BH/\eta
\sqrt{\mu_0 \rho}$.
\begin{figure}
\psfig{figure=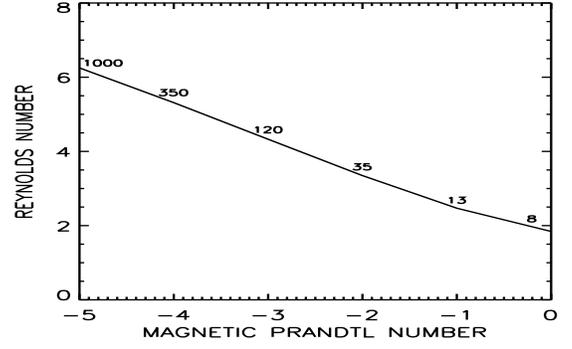,width=8.8cm,height=5.0cm}
     \caption{The main results for $\hat \eta=0.5$ and $\hat \mu = 0.33$: The 
     critical Reynolds numbers for given magnetic Prandtl numbers 
marked with those Hartmann numbers where the Reynolds number is minimal.}
\label{fig5}
\end{figure}

\subsubsection{Wide gap}
Let us now vary the  size of the gap. In view of the experimental
possibilities, we shall only work for conducting fluids  with the magnetic
Prandtl number of sodium, i.e. $10^{-5}$. In  the present Section 
cylinders with a gap with $\hat \eta=0.25$ are discussed. The outer cylinder is
either resting (Fig. \ref{fig7}) or it is rotating with a frequency fulfilling
the Rayleigh criterion for stability (Fig. \ref{fig8}). In the first case, of 
course, there is a solution without magnetic field, i.e. for ${\rm Ha}=0$. The 
corresponding Reynolds number is 79.  Note again that  a minimum appears 
for Pm=1 which, however, does not survive the decrease of the magnetic
Prandtl number to realistic small values.

The minimum always exists, however, for experiments with a rotating outer 
cylinder, e.g. for $\hat \mu=0.1$ (Fig. \ref{fig8}). The resulting critical
Reynolds number  is $1.15 \cdot 10^6$ and the critical Hartmann number is
about 500. Let us turn to first estimates.
With $\nu = 10^{-2}$ cm$^2$/s 
the frequency $f$ of the inner cylinder is 
\beg
f={1.6 \cdot 10^{-5} {\rm Re} \over \hat\eta (1-\hat\eta)}
\left({10\ {\rm cm} \over R_{\rm out}}\right)^2 \ \  {\rm Hz} ,
\label{f}
\ende
so that here
\beg
f= {98 \over \left(R_{\rm out}/10\ {\rm cm}\right)^2} 
\; {\rm Hz}, 
\label{f85}
\ende
 corresponding to the frequency of  about 16 Hz for
a container  with an outer  radius of  25 cm.\footnote{very close to the
parameters of the experiments in \cite{LFS}}

For the Hartmann number
with the density for liquid sodium ($\rho \simeq 1\ {\rm g/cm}^3$)
one finds
\beg
{\rm Ha} = 282 \left({B\over {\rm Gauss}}\right) \left({R_{\rm out} 
\over 10\ {\rm cm}}\right) \sqrt {\hat\eta(1-\hat\eta){\rm Pm}},
\label{Ha}
\ende
hence for $\hat \eta=0.25$ and Pm=$10^{-5}$,
\beg
{\rm Ha}= 0.39\  \left({B\over {\rm Gauss}}\right) \left({R_{\rm out}
 \over 10\ {\rm cm}}\right)
\label{Ha1}
\ende
results. For a container of (say) 25 cm a field of  500 Gauss  yields thus a
Hartmann number of 500. 
Note that this result has only a weak dependence on $\hat
\eta$.  It is thus not a problem to reach Hartmann numbers of order
$10^4$ with the standard laboratory equipment.

We have to realize that Fig. \ref{fig7} only displays suppression of the 
instability by the magnetic field for Pm=10$^{-5}$. There is no minimum 
of the Reynolds number due to the MRI instability.
This effect is a consequence of the low magnetic Prandtl number.
As it  must, the instability disappears for ${\rm Ha}=0$ and $\hat\eta = 0.25$
if $\hat\mu=0.1$ (Fig. \ref{fig8}). But here we find the instability again
for a finite Hartmann number. For ${\rm Ha} \approx 500$ an instability
occurs for a Reynolds number of about $10^6$. 
An experiment with a (say) Reynolds number of $1.5\cdot 10^6$ 
and an increasing magnetic field should yield the MRI instability between
two known very sharp limits\footnote{....if not a nonlinear hydrodynamic 
instability exists for the given (high) Reynolds number \cite{RZ}}.
The rotation frequency of the inner cylinder must fulfill the above relation
(\ref{f85}), i.e. a container with an outer radius of 31 cm must rotate with
a frequency of 10 Hz (see \cite{LFS}).
\begin{figure}
\psfig{figure=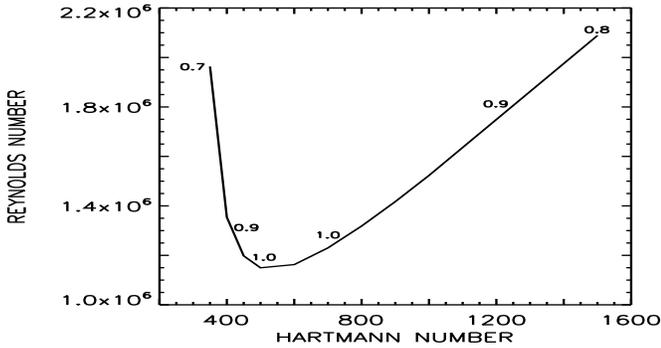,width=8.8cm,height=5.0cm} 
     \caption{Wide gap ($\hat\eta=0.25$): The same  as in Fig. \ref{fig7} 
     (Pm$=10^{-5}$)
      but for a rotating outer cylinder with $\hat\mu=0.1$.} 
\label{fig8}
\end{figure}

\subsubsection{Small gap}
For small gaps and  resting outer cylinder  there is no minimum due to MRI for 
 magnetic Prandtl numbers equal or smaller than 1  
 (see Figs. \ref{fig10} and \ref{fig11}) but it exists for e.g. Pm=10 (not 
 shown). 
If  the outer cylinder starts to rotate then the hydrodynamic instability
goes to infinity and a minimum again appears due to the MRI (Fig. \ref{fig12}).
However, the Reynolds numbers are much too high for a technical realization
(inner rotation frequency is of order $10^3$ Hz). 
Obviously, MHD Taylor-Couette flows with too  small gaps between the cylinders 
are not suitable for experimental work. 
\begin{figure}
\psfig{figure=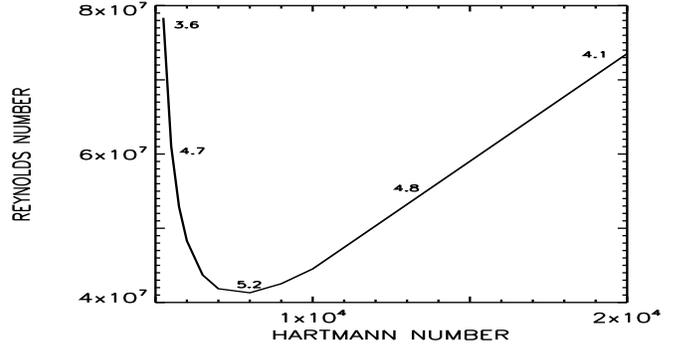,width=8.8cm,height=5.0cm} 
     \caption{Small  gap ($\hat\eta=0.95$): The same as in Fig. \ref{fig11} but
      for a rotating outer cylinder with $\hat\mu=0.95$. Pm =10$^{-5}$, the 
      critical Reynolds number is extremely high.} 
\label{fig12}
\end{figure}
\section{Results for isolating walls}
Containers with isolating walls must be considered. The (complicated) boundary
conditions are then given by the relations (\ref{nonin}) and (\ref{nonout}).
Surprisingly, the basic differences  can already be 
demonstrated by the simplest model given in Fig. \ref{fig11a} for 
resting outer cylinder and Pm=1 (see Fig. \ref{fig} for comparison). 
Of course, the profiles start for Ha=0 with the same Reynolds number.   
The minimum, however, is deeper  than in  Fig. \ref{fig} and the 
corresponding Hartmann number is higher. Note that the 
vertical wavelength in the minimum is {\em larger} than in containers 
with conducting walls. 
\begin{figure}
\psfig{figure=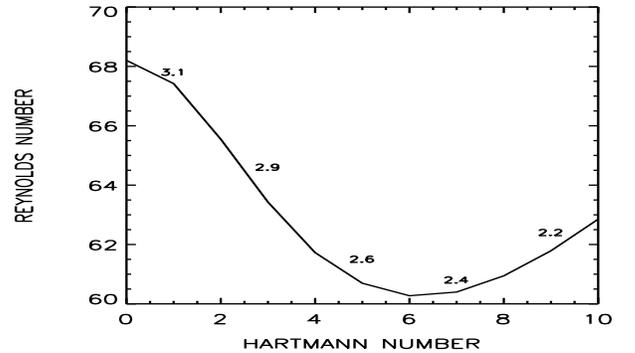,width=8.8cm,height=5.0cm} 
    \caption{
      The same as in Fig. \ref{fig} but for isolating walls.
} 
\label{fig11a}
\end{figure}
We shall check these findings in the following under restriction to
a small magnetic Prandtl number (10$^{-5}$) and for rotating outer cylinders 
for small (Fig. \ref{fig17}), medium (Fig. \ref{fig14}) and wide 
(Fig. \ref{fig16}) gaps.  The results must be compared
with the results given in Figs. \ref{fig4}, \ref{fig8} and \ref{fig12} valid
for conducting walls. For small and for medium gaps
one finds indeed that i) the minimal Reynolds numbers are smaller,
ii) the corresponding Hartmann number is higher and
iii)  vertical wave number is smaller
(i.e. the cells of Taylor vortices are vertically more elongated)
for the container with isolating walls.
For wide gaps the critical Reynolds
number is slightly higher for the container with nonconducting walls now 
but the vertical size of the cell is the same.  
\begin{figure}
\psfig{figure=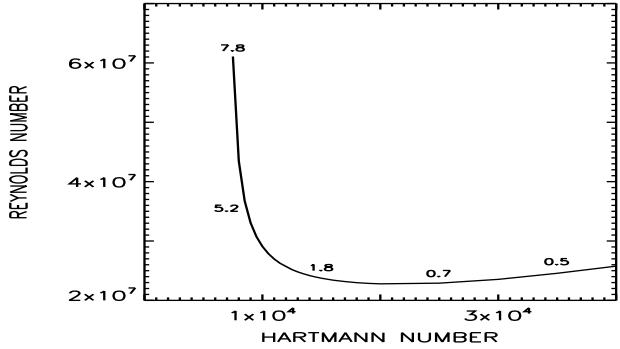,width=8.8cm,height=5.0cm} 
    \caption{Small gap ($\hat \eta = 0.95$): The same as in Fig.
    \ref{fig12} but for the rotating outer cylinder ($\hat \mu =0.95$)
     embedded in vacuum. Pm=10$^{-5}$.} 
\label{fig17}
\end{figure}
\begin{figure}
\psfig{figure=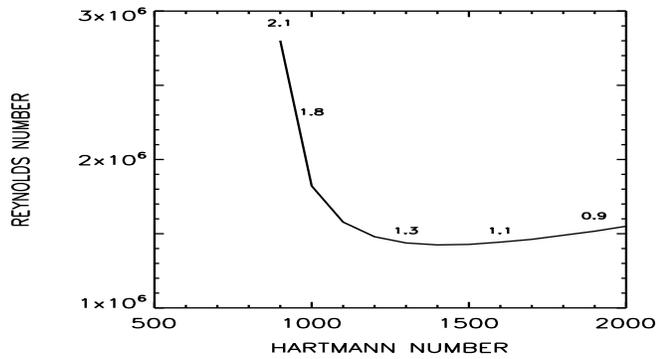,width=8.8cm,height=5.0cm} 
    \caption{Medium-size gap ($\hat \eta =0.5$): The same as in Fig.
    \ref{fig4} but for the rotating outer cylinder ($\hat\mu = 0.33$) 
    embedded in vacuum. Pm$=10^{-5}$.} 
\label{fig14}
\end{figure}
\begin{figure}
\psfig{figure=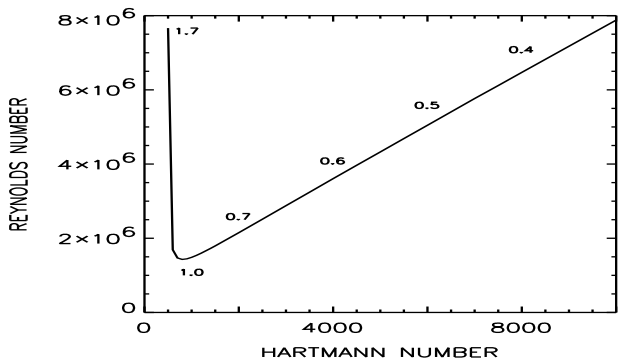,width=8.8cm,height=5.0cm} 
    \caption{Wide gap ($\hat\eta =0.25$): The same as in Fig. \ref{fig8}
    but for the  rotating outer cylinder ($\hat\mu=0.1$) embedded in
    vacuum. Pm=10$^{-5}$.} 
\label{fig16}
\end{figure}
\section{Vertical cell structure}
 The unstable Taylor-Couette flow forms Taylor  vortices.  
With our normalizations the vertical extend $\delta z$ of a Taylor vortex is 
given by
\beg
{\delta z \over R_{\rm out} - R_{\rm in}} = {\pi \over k}
\sqrt{{\hat \eta \over 1-\hat\eta}}.
\label{delz}
\ende
The dimensionless vertical wavenumber $k$ is given in all  the above 
figures.
 
In the case of hydrodynamically unstable flows
we have $\delta z \simeq R_{\rm out} - R_{\rm in}$ for small magnetic field
(Ha$\simeq0$) independently on gap size and boundary conditions 
(see Figs. \ref{fig}, \ref{fig10}, \ref{fig7}, \ref{fig11a}).
The cell has the same vertical extend as it has in radius (see \cite{Ko}).

As  all our figures demonstrate
the  influence of strong magnetic fields on turbulence consists on suppression
and deformation. 
The deformation consists on a prolongation of the cell structure in vertical 
direction (\cite {R}) so that    
$\delta z$ is expected to become larger and larger (the wave number
becomes smaller and smaller) for increasing
magnetic field. It is indeed true  for  Pm$\sim$1, but for  smaller  Pm
the vertical cell size has a minimum for an intermediate value of the magnetic
field (see Figs. \ref{fig0}, \ref{fig11}, \ref{fig7}).

The cell size is minimal for the critical Reynolds number for all calculated
examples for hydrodynamically stable flow and conducting boundary (see e.g.
Figs. \ref{fig2}, \ref{fig8} and \ref{fig12}).  This  is not true, however,
for containers with isolating walls  for which the  cell size  grows with
increasing magnetic field. 
For experiments with the critical Reynolds numbers the vertical cell
size is generally 2...3 times larger than the radial one.
The dependence of the vertical cell size on the magnetic Prandtl number
is illustrated by the Fig. \ref{fig6}. The smaller the magnetic Prandtl
number the bigger are the cells in  vertical direction.

The influence of boundary conditions on the cell size disappears for wide gaps
 between the cylinders. For the small and medium gap, however,  one finds 
 the cells  vertically more
elongated for  containers with isolating walls.
\begin{figure}
\psfig{figure=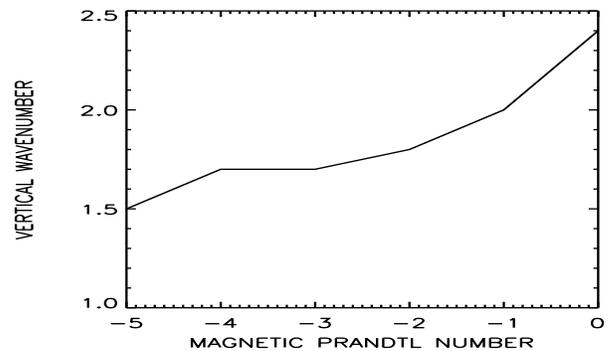,width=8.8cm,height=5.0cm} 
     \caption{The same as in Fig. \ref{fig5} but for the  vertical wave number} 
\label{fig6}
\end{figure}
\section{Discussion}
We have shown how the MRI works in Taylor-Couette flow 
 experiments
for fluids with  high and low electrical conductivity and for conducting walls
as well as  for isolating ones.  For given microscopic viscosity
the electrical conductivity determines the magnetic Prandtl number which in the
present paper is varied  between 1 and $10^{-5}$.

For Pm=1 and large enough gap between the cylinders the MRI is realized by a
clear minimum of the Reynolds number for certain 
(critical) magnetic fields with Hartmann numbers of order 10. The existence of
the minimum does not strongly depend on the rotation rate of the outer cylinder
 -- provided it rotates slower than the inner cylinder
(Figs. \ref{fig}, \ref{fig2} and also \ref{fig11a}).     

There are drastic differences, however, for small magnetic Prandtl numbers
 or small gap between the cylinders. 
The minima completely disappear for resting outer cylinders
(see Figs. \ref{fig0}, \ref{fig10} and \ref{fig11}). But they survive for
rotating outer cylinder (see Figs. \ref{fig2}....\ref{fig4} and also
\ref{fig14}). 

The coordinates of the minima strongly depend  on the magnetic Prandtl number
Pm. The critical Reynolds number scales as 1/Pm with the magnetic Prandtl
number and the critical Hartmann number scales as $1/\sqrt{{\rm Pm}}$ 
 for small Pm (see Fig. \ref{fig5}). We find the surprising result therefore
that for sufficiently small magnetic Prandtl number both the {\em magnetic}
Reynolds number $C_\Omega$ and the {\em magnetic} Hartmann number
Ha$^*$ (defined after (\ref{COm}) and (\ref{Ha-st})) 
 depend only weakly on the magnetic Prandtl number. 

Generally, the presented results tend to reduced critical Reynolds numbers for
isolating  rather than conducting walls. The power-law exponent (-1) which 
here results for an infinite cylinder is stronger than the value (-0.65)
which has been found for a finite cylinder with pseudo-vacuum boundary
conditions and a aspect ratio of  10  (cf. \cite{RZhang}). 

>From Eq. (\ref{f}) with $\nu=10^{-2}$ cm$^2$/s, $\hat \eta$=0.5 and Re$\simeq
1.8 \cdot 10^6$ (see Fig. \ref{fig5}) for Pm$=10^{-5}$ follows
\beg 
f= {115\over \left(R_{\rm out}/10\ {\rm cm}\right)^2} \ {\rm Hz}
\label{fff}
\ende
for the frequency of the inner cylinder. Hence, a container with  an outer
radius of 30 cm and an inner radius of 15 cm 
requires a rotation of about 10 Hz in order to exhibit the MRI for liquid
sodium with its magnetic Prandtl number of 10$^{-5}$. After (\ref{HA})
the required magnetic field is about 900 Gauss.

 The MRI is considered here only for axisymmetric
disturbances. According to small gap small Pm results \cite{CC98},
the non-axisymmetric disturbances can be more unstable for small
magnetic field. We are going to consider the influence
of non-axisymmetric disturbances on MRI in a forthcoming paper.


\end{document}